\documentclass{aa}
\usepackage{txfonts}
\usepackage{graphicx}

\def\ie{{i.e.}}

\def\etal{et al.}
\def\*{{$^*$}}
\usepackage{natbib}
\usepackage{comment}

\begin{document}

\title{Statistical properties of giant pulses from the Crab pulsar}

\author{M.V. Popov\inst{1}
	\and B.Stappers\inst{2}
       }

\offprints{M.V. Popov}
\mail{mpopov@asc.rssi.ru}
\institute{Astro Space Center of the Lebedev Physical Institute,
           Profsoyuznaya 84/32, Moscow, 117997 Russia
           \and
	   Astronomical Institute "Anton Pannekoek", University of Amsterdam,
	   Kruislaan 403, 1098 SJ Amsterdam,The Netherlands; Stichting ASTRON,
	   Postbus 2, 7990 AA, Dwingeloo, The Netherlands           
           }

\abstract{}
{We have studied the statistics of giant pulses from the Crab pulsar  for
the first time with particular reference to their widths.}
{We have analyzed data collected during
3.5~hours of observations conducted with the Westerbork Synthesis
Radio Telescope operated in a tied-array mode at a frequency of 1200~\mbox{MHz}.
The PuMa pulsar backend provided voltage recording of X and
Y linear polarization states in two conjugate 10~\mbox{MHz} bands.
We restricted the time resolution  to 4~$\mu s$ to match the scattering on the
interstellar inhomogeneities.}
{In total about 18000 giant pulses (GP)
were detected in full intensity with a threshold level of
6$\sigma$. We analyzed cumulative probability distribution (CPD) of giant pulse
energies for groups of GPs with different effective
widths in the range 4 to 65~$\mu s$.  The CPDs were found to manifest
notable differences for the different GP width groups. The slope of a
power-law fit to the high-energy portion of the CPD evolves from $-$1.7
to $-$3.2 when going from the shortest to the longest GPs. There are
breaks in the CPD power-law fits indicating flattening at low energies
with indices varying from $-$1.0 to $-$1.9 for the short and long GPs,
respectively. The GPs with a stronger peak flux density were found to
be of shorter duration. We compare our results with
previously published data and discuss the importance of these
peculiarities in the statistical properties of GPs for the theoretical
understanding of the emission mechanism responsible for GP generation.}
{}

\keywords{pulsars: general~-- {\bf pulsars: individual: B0531+21~}--
          Methods: statistical~--
          Radiation mechanisms: non-termal}

\date{Received ~~~~~~~; accepted ~~~~~~~}

\titlerunning{Statistical Properties of Giant Pulses}
\authorrunning{M.V. Popov \& B. Stappers}
\maketitle

\section{\bf Introduction}
\label{intro}

Giant pulses are one of the most striking phenomena of pulsar radio
emission. Their flux density can exceed thousands of times the average
pulse-integrated flux. Although recently there are reports of 
detecting excessively strong pulses from a number of pulsars
\citep{johnston2001, kramer2002, romani2001, johnston2002, kuzmin2004,
  ershov2003, knight2006}, only the Crab pulsar and the millisecond
pulsar B1937+21 generate giant pulses numerous enough to study
their statistical properties. The very large fluxes of giant pulses
are coupled with an extremely short duration. Indeed, the overwhelming
majority of giant pulses from the millisecond pulsar B1937+21 are
shorter than 15 ns \citep{soglasnov2004}, while giant pulses from the
Crab pulsar have a mean width of about few microseconds
\citep{hankins2000} with occasional bursts shorter than 2 ns
\citep{hankins2003}. \\
The longitude position of giant pulses is
remarkable in that it coincides with the position of high-energy
emission \citep{moffett1996, cusumano2003, johnston2004, knight2006}.
Furthermore, giant pulses originate in a very narrow phase window that
in general does not correspond to the phase window of regular radio
emission. In the millisecond pulsar B1937+21, giant pulses are observed
at the very trailing edge of the average profile in both the main
pulse and the interpulse \citep{cognard1996, kink2000,
  soglasnov2004}. \citet{popov2006} has recently  suggested that
giant pulses from the Crab pulsar are also seen at the trailing
edge of the regular radio emission window, which they consider to be the
precursor. In other words, they consider that radio emission in the
main pulse and the interpulse of the Crab pulsar consists completely
of giant pulses.  Lastly, an essential property of giant pulses, by
which they may be distinguished from normal strong pulses, is the
distribution of their flux  densities, which appears to follow a power law in
contrast with the Gaussian or exponential flux distribution typical
of regular (ordinary) individual pulses \citep{backer1971, hess1974,
  ritchings1976}.  In this paper we present an analysis of giant
pulse-energy distribution based on the observations conducted with the
Westerbork Synthesis Radio Telescope (WSRT) at 1200 \mbox{MHz} using the PuMa
\citep{voute2002} pulsar backend.

\section{Observations and data reduction}
\label{obs}
The observations were carried out in November 2003 as part of a
multi-frequency observing campaign that also included Jodrell Bank at
1420~\mbox{MHz}, Effelsberg at 8350~\mbox{MHz}, Kalyazin at 600~\mbox{MHz}, Pushchino at
111~\mbox{MHz}, and Kharkov (UTR-2) at 23~\mbox{MHz}. Simultaneous optical
observations were also made with the 6-m telescope of the Special
Astrophysical Observatory and 2.8-m telescope at La Palma. The MAGIC
and HESS gamma-ray telescopes in La Palma and Namibiya also
participated.  While some results obtained at the separate
observatories have already been published \citep{jessner2005, popov2006},
the joint analysis of multi-frequency observations will be presented
in future publications.\\
We have analyzed the data
obtained in observations with the WSRT during about 3.5 hours in two
conjugate bands of 10~\mbox{MHz} each at a central frequency of 1197~\mbox{MHz}.
Baseband voltages of X and Y linear polarization states were recorded
with two-bit sampling at Nyquist frequency. The coherent predetection
dedispersion technique originally developed by Hankins
\citep{hankins1971, hankins1975} was used to remove dispersion
smearing in the received pulsar signal.  We took the value of
dispersion measure (56.757~$\mathrm{pc~cm^{-3}}$) and the timing model from the
Jodrell Bank Monthly Ephemeris \citep{lyne1982}. The technique
provides a formal time resolution of 100 ns, but the expected
pulse-broadening time due to interstellar scintillations is in the
range 2 to 4~$\mu s$, based on the estimation made by
\citet{kuzmin2002}.  Therefore, we averaged the recorded signal after
dispersion removal and square-low detection synchronously with
a topocentric pulsar period into 8192 bins per period and with a resulting
sampling interval close to 4.1~$\mu s$. The total intensity time
series was finally formed and used for detection of giant pulses with
the threshold level of 6$\sigma$.  To provide better sensitivity for
wider giant pulses, we developed a searching procedure which 
progressively tried to increase the averaging time interval ($\tau_i$) by 1, 2,
3, 4, 6, 8, 12, 16, 24, 32, and 48 samples, \ie from 4.1~$\mu s$ to
196~$\mu s$. The averaging time with the best signal-to-noise ratio
(SNR) was selected as $W_e$ for the GP, detected simultaneously at several
averaging times. The pulse energy $E$ (or average flux density) was
calculated for the GP as a product of the SNR and the pulse width,
equal to the averaging time, which corresponds to the best SNR value. In
this approach we have different threshold levels both in peak flux
density $F_p$ and in pulse energy $E$ for each separate averaging
time. The value of the root-mean-square deviation (RMS or $\sigma$)
goes down with increasing time averaging as the square root of
time. Therefore, the threshold level in peak flux density $F_p$
decreases with increasing averaging time as $F_p(\tau)\propto 1/\sqrt
{\tau}$, while the threshold level in pulse energy $E$ increases with
$\tau$ as $E(\tau)\propto \sqrt {\tau}$.\\
For the Crab pulsar, the
system temperature is notably influenced by the impact of the Crab
Nebula, whose flux density can be approximated by the
relation $F_\nu=955\nu^{-0.27}$~\mbox{Jy} ($\nu$ in GHz) \citep{allen,
bietenholz1997}, which gives $F_{1200}=909$~\mbox{Jy} at 1200 \mbox{MHz}.  In our
observations we used all 14 dishes of the WSRT in tied-array mode,
where all telescopes are added coherently, with the width of
synthesized beam being equal to 34 arcsec, thus reducing the contribution
from the Crab Nebula to the system noise by a factor
$f_\nu=\Omega_A/\Omega_{CN}=0.14$ to the value
$F_\mathrm{CN}=127.6$~\mbox{Jy}. Here $\Omega_\mathrm{CN}$ is a solid angle
of the Crab Nebula, and $\Omega_\mathrm{A}$
a solid angle of the intersection
of the synthesized beam with the Crab Nebula.
However, there is still the contribution from the
individual dishes; the system temperature of every 25-m dish was
increased by the Crab Nebula emission by $\Delta
T_\mathrm{CN}=F_\mathrm{1200}G=86.3$~\mbox{K}, where G is the gain of a single telescope,
equal to 0.095 at 1200~\mbox{MHz}. The intrinsic system temperature in the
absence of the Crab Nebula is 30~\mbox{K} at 1200~\mbox{MHz}, and the resulting
total system temperature is thus 116.3~\mbox{K}, equivalent to
$F_\mathrm{sys}=87.4$~\mbox{Jy} for the
gain in the tied array (1.33~\mbox{K/Jy}), which
when combined with the nebula contribution to the tied array becomes
$F_\mathrm{tot}=F_\mathrm{sys}+F_\mathrm{CN}=215$~\mbox{Jy}.
One can see that WSRT reduces the
contribution of the Crab Nebula to the system noise considerably when
compared with the single dish observations at this frequency. This
improvement will permit us to follow the statistics of giant pulses
from the Crab pulsar to the lower energies; namely, the limiting RMS
($\sigma$) value in one polarization is $F_\mathrm{tot}/\sqrt{B\tau}$, with
$B=10$~\mbox{MHz}, and $\tau=4.1~\mu s$ the RMS is equal to 33.5~\mbox{Jy}.
For total intensity, the RMS will be reduced by $\sqrt{2}$, and the
threshold of $6\sigma$ in peak flux density will be equal to 142~\mbox{Jy}.
The threshold will be lower for averaged records, with a limiting value
of 20~\mbox{Jy} for giant pulses, which have a maximum effective width of about
200~$\mu s$.\\
In total 17869 giant pulses were detected over 370000
pulse periods with 14994 giant pulses located at the longitude of the
main pulse and 2875 giant pulses associated with the interpulse. The
background rate was determined by counting events at quiescent phases
of the pulsar period and the corresponding values were taken into
account in our statistical calculations.  In fact, at the lowest
range, between $6\sigma$ and $7\sigma$, only about $10\%$ of all pulses
we potentially incorrectly identified as giant pulses.\\

\section{Results}
\label{res}
\begin{figure}

\includegraphics[height=9cm,width=8cm]{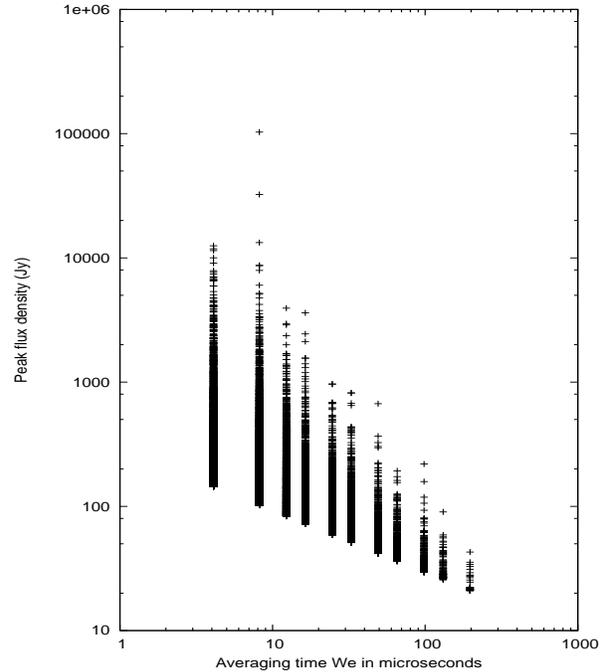}

\vskip 2mm

\caption{Peak flux density versus the effective puse width $W_e$ for all
detected giant pulses.}
\label{FluxWidth}
\end{figure}
\begin{figure*}[htp]

\includegraphics[height=6.2cm,width=6.5cm]{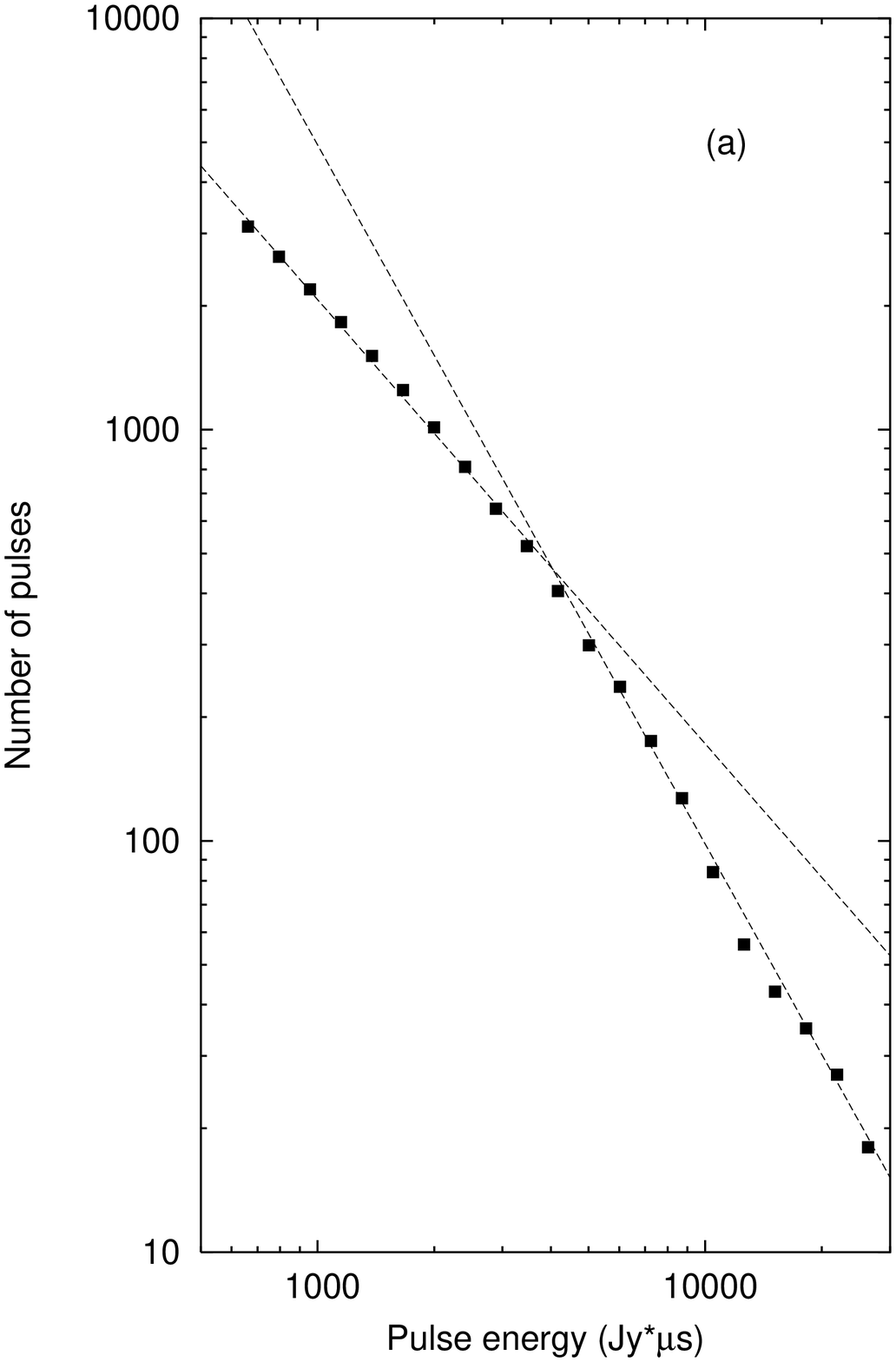}
\hspace{1cm}
\includegraphics[height=6.2cm,width=6.5cm]{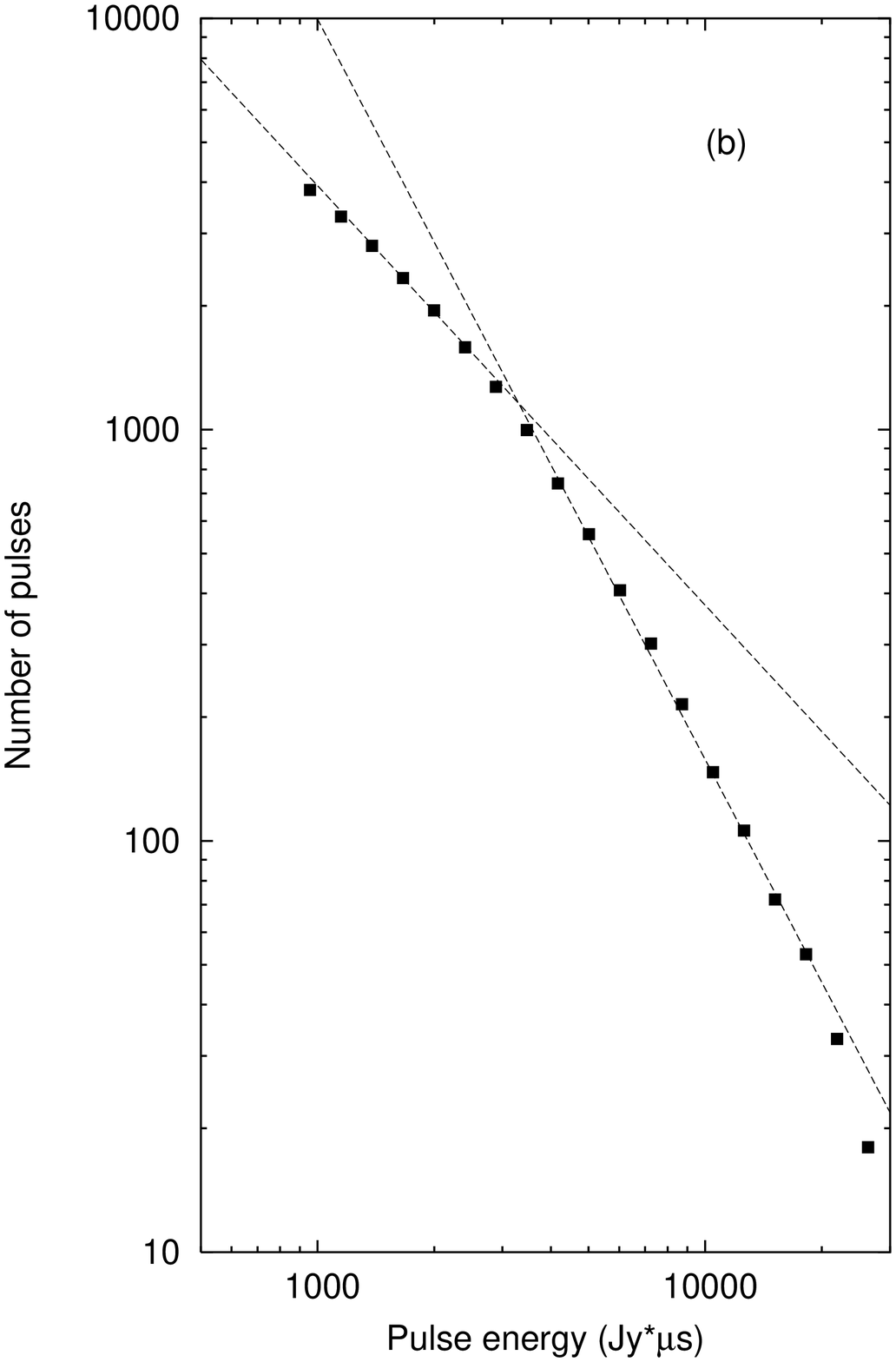}

\includegraphics[height=6.2cm,width=6.5cm]{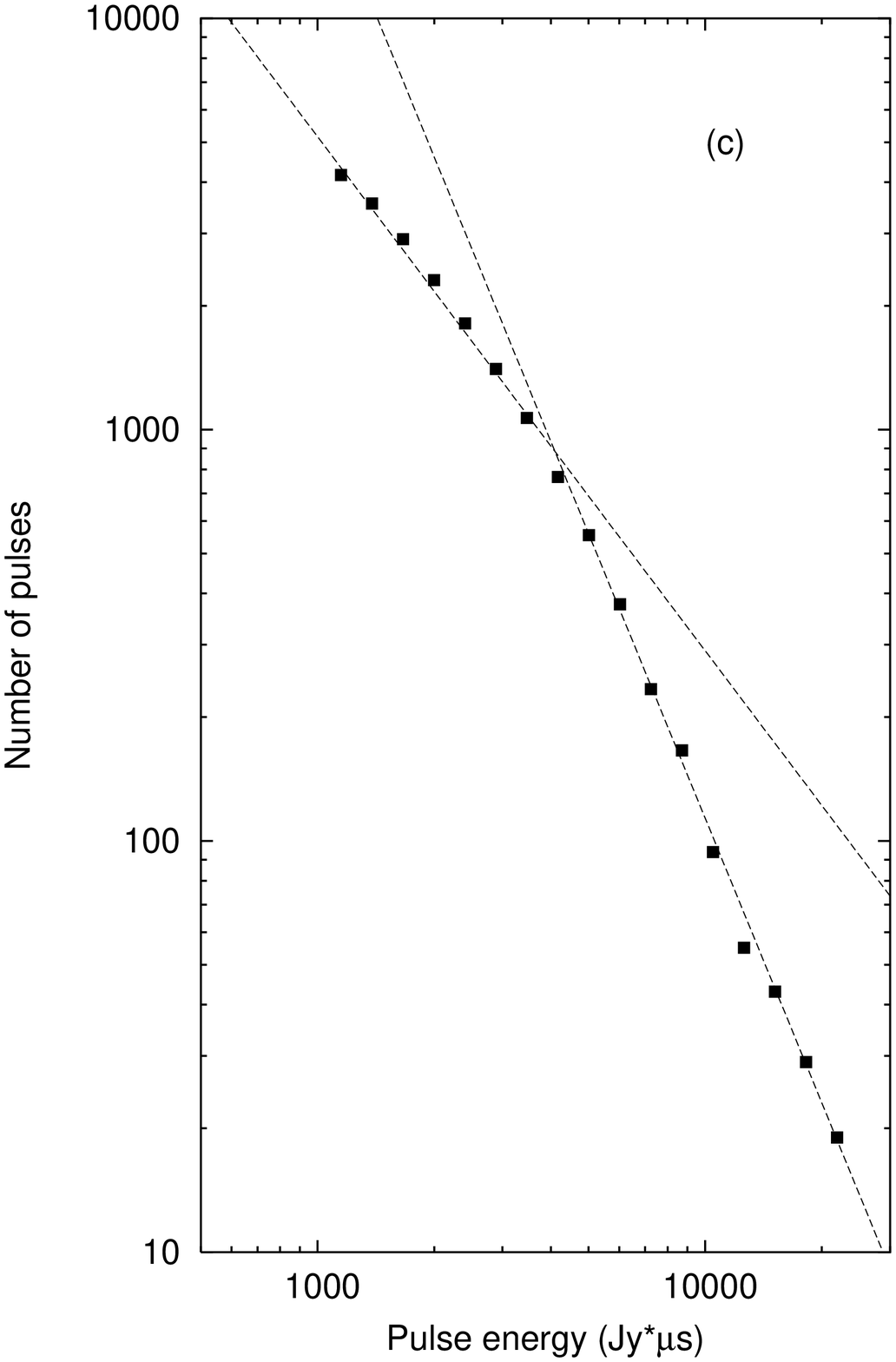}
\hspace{1cm}
\includegraphics[height=6.2cm,width=6.5cm]{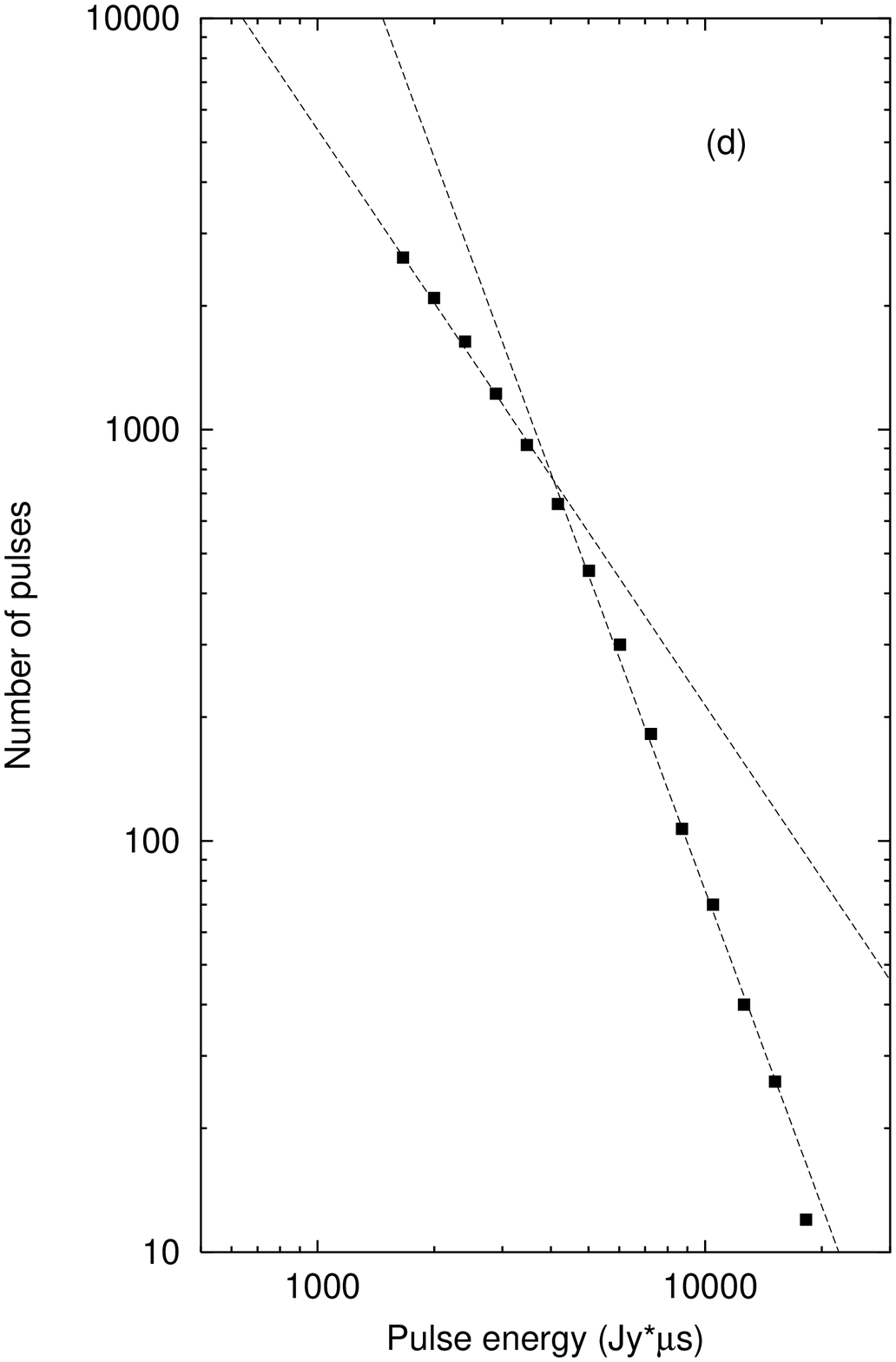}

\includegraphics[height=6.2cm,width=6.5cm]{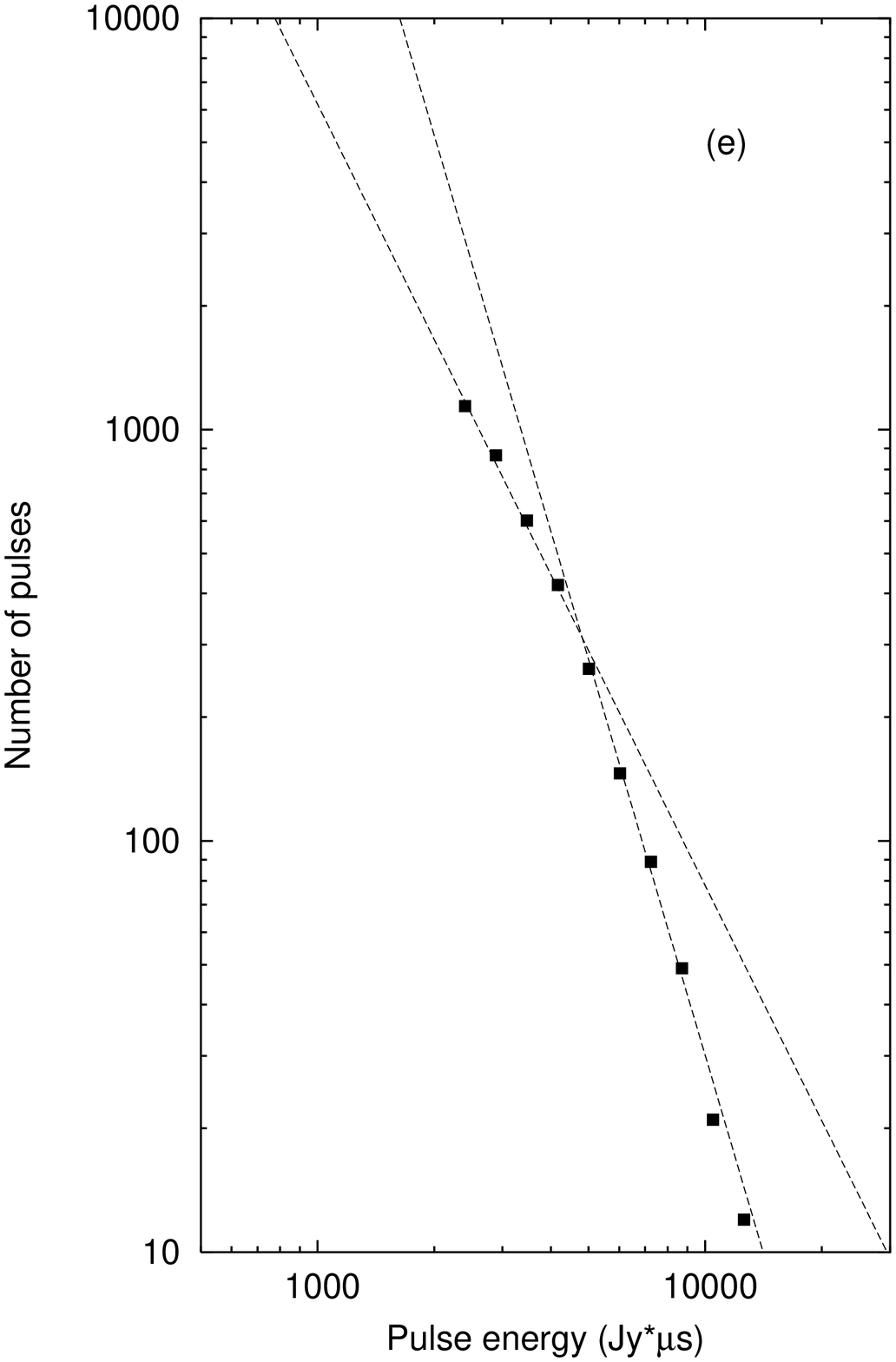}
\hspace{1cm}
\includegraphics[height=6.2cm,width=6.5cm]{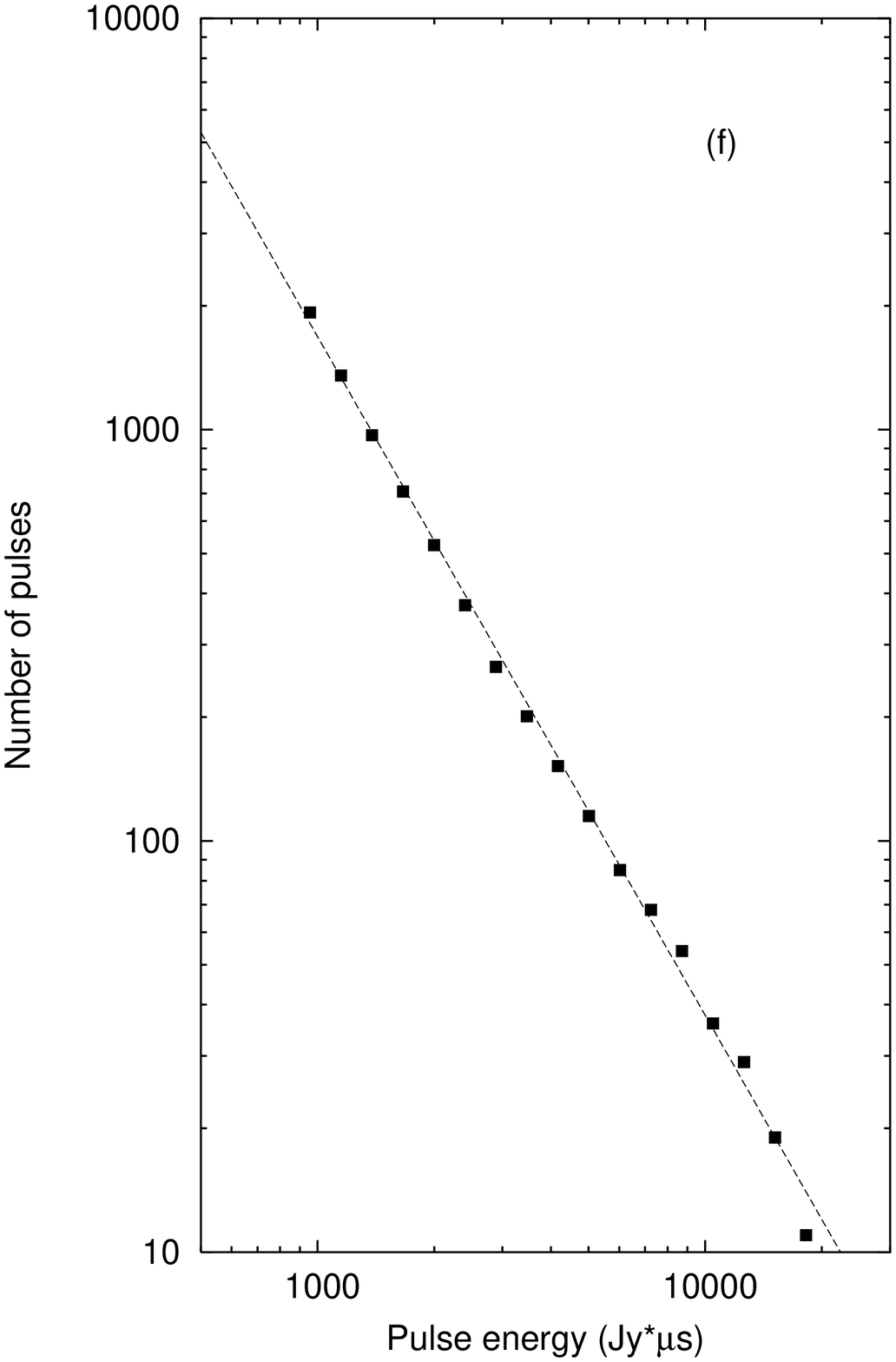}

\vskip 2mm

\caption{The CPD of giant pulses with
  pulse energy above the displayed value for different groups of giant
  pulses classified by their effective width $W_e$ as described in the
  Table \ref{powerfit}.  Straight lines represent the fit by power-law
  functions with the parameters indicated in the Table \ref{powerfit}.
  The last plot (f) represents giant pulses belonging to the
  interpulse.}

\label{CumProb}

\end{figure*}

Figure \ref{FluxWidth} represents the complete set of data showing
 peak flux density $F_p$ versus the effective pulse width $W_e$.
The striped nature of the diagram is caused by the procedure of giant
pulse detection, which is based on the discrete values of averaging
time $W_e$. The diagram demonstrates the peculiarity of the statistics
on giant pulses: the strongest pulses
clearly have a shorter duration; namely, there are no
pulses with peak flux density greater than 1000~\mbox{Jy}  that are wider than 16~$\mu
s$. As a consequence, giant pulses with different durations have
different distributions in pulse energy, and the distribution in giant
pulse width depends on the range of pulse energy used.  Therefore, in
our analysis, we separated the giant pulses belonging to the main
pulse longitudes into five groups by their widths: (a) -- GPs with the
effective width $W_e$ of 4.1~$\mu s$, (b) -- GPs with the $W_e$ of
8.2~$\mu s$, (c) -- combined group of GPs with the $W_e$ of 12.3 and
16.4~$\mu s$, (d) -- combined group of GPs with the $W_e$ of 24.5 and
32.8~$\mu s$, and (e) -- combined group of GPs with the $W_e$ of 49.2
and 65.6~$\mu s$.  A separate group (f) was formed for the GPs
detected at the longitude of the interpulse with the $W_e$ 4.1, 8.2,
and 12.3~$\mu s$. Then, pulse energy, or
integrated flux density, was calculated as $E=F_p\times W_e$.\\
In our analysis we use cumulative probability distribution (CPD),
which gives the number of pulses $N(E)$, above pulse energy
$E$. Some authors prefer to calculate the probability distribution (PD)
by giving the number of pulses $n(E)$ per interval of energy $dE$. If
a power-law fit is used for the PD function as $n(E)\propto E^{-\beta}$,
then a power-law fit will be also valid for the CPD function
$N(E)\propto E^{-\gamma}$, since
$$N(E>E_y)=\int\limits_{E_y}^\infty n(E)dE \propto E_y^{-\beta+1}$$
with $\gamma=\beta-1$ in the absolute value.\\
Figure \ref{CumProb} displays the CPD of pulse energies
for all the aforementioned groups of GPs.  The CPD
for the GPs belonging to the interpulse can be fitted with a power-law
function with $\gamma=1.6\pm0.1$.  The CPDs for the GPs in the main
pulse manifest a break in the power-law index at certain values of pulse
energy, which is slightly different for the different groups. Table
\ref{powerfit} contains the results of a least-square solution for
the power-law fit.
The first column indicates the group of GPs, $W_e$ is the averaging time for
the group in $\mu \mathrm{s}$, $\gamma_\mathrm{high}$ and $\gamma_\mathrm{low}$
are indices
of the power law function for the high energy tail and low energy
portion of the CPD, respectively, and $E_\mathrm{break}$ indicates the pulse
energy where the slope of the power law changes.
The break energy values were obtained as the crossing
point between two straight lines defined by the least-square solutions.
Relative inaccuracy is within 10\% at  the RMS level.\\

\medskip
\begin{table}
\caption{The list of parameters of the power-law fits to the
  CPDs displayed in Fig.2.}
\label{powerfit}
\begin{tabular}{ccccc}
\hline
\hline
Group&$W_\mathrm{e} (\mu s)$&$\gamma_\mathrm{high}$&$\gamma_\mathrm{low}$&$E_\mathrm{break}(\mathrm{Jy}\cdot\mu \mathrm{s})$\\
\hline
(a) & 4.1 & 1.7 & 1.1 & 4000\\
(b) & 8.2 & 1.8 & 1.0 & 3500\\
(c) & 12.3, 16.4 & 2.3 & 1.2& 4000\\
(d) & 24.5, 32.8 & 2.5 & 1.4& 4000\\
(e) & 49.2, 65.6 & 3.2 & 1.9& 5000\\
(f) & 4.1, 8.2, 12.3& 1.6 & 1.6&  \\
\hline
 
\end{tabular}
\end{table}

The absence of any break in the power-law fit for the CPD of the GPs
belonging to the interpulse longitudes is good proof that the
observed breaks in the CPDs of the main pulse GPs were not caused by
selection effects, since we used exactly the same procedure for GPs
detection irrespective of pulse longitude. A possible explanation is that we
observe the regions where the GPs are generated in the the main pulse
and the interpulse at different impact angles relative to the GP beam,
which has a half width of about 5 degrees. Indeed, the comparison of
the CPDs (combined for the width groups (a), (b), and (c)) for the
main pulse and the interpulse shows that the CPDs would be identical
if one were to multiply the interpulse GP energies by a factor of
about 4 and increase their rate of occurrence by a factor of about 2.
Therefore, the weakest detected interpulse GPs with a pulse energy of
about 1000~$\mathrm{Jy}\cdot\mu \mathrm{s}$, when observed at the same impact angle as
the main pulse GPs (multiplied by 4), will have pulse energies of
4000~$\mathrm{Jy}\cdot\mu \mathrm{s}$, which is above the break point.

\section{Discussion and conclusions}
\label{disc}
The CPD function of the GPs
originating at the longitudes of the main pulse manifests gradual
changes in the power-law index from $-$1.7 to $-$3.2 at the high energy part
of the CPD, and from $-$1.0 to $-$1.9 at the low energy portion of the CPD. As
a rough guide, all GPs can be separated into two main groups: short GPs
belonging to groups (a) and (b), and long GPs with an $W_e$ greater
than 8.1~$\mu s$. It is interesting to note that the CPD power-law
index of about $-$1.7 for short GPs is close to the exponent
determined for the CPD of the GPs from the millisecond pulsar
B1937+21, which was found to be equal to $-$1.8 at 430~\mbox{MHz}
\citep{cognard1996}. The value was confirmed by \citet{kink2000}, and
recently estimated by \citet{soglasnov2004} as $-1.4$ at 1650~\mbox{MHz}.
All GPs from the millisecond pulsar are very short, lasting only
1-2~$\mu s$ as measured by \citet{kink2000}, while
\citet{soglasnov2004} have found that the majority of GPs from this pulsar
are shorter than 15~ns.\\
\citet{argyle1972} were the first to present the CPD for GPs from the
Crab pulsar. They combined the results of observations made at 146 \mbox{MHz}
with the 26-m Penticton radio telescope and with the 46-m dish at the
Algonquin Radio Observatory and found that the CPD was
consistent with the power-law exponent of $-$2.5 for the main pulse and
with the exponent for the interpulse events equal to $-$2.8.\\
Recently \citet{cordes2004} in their multifrequency study of the Crab
pulsar's giant pulses presented histograms of GP peak amplitudes (S)
at 0.43 and 8.8~\mbox{GHz}.  The histograms represent the 
PD functions, i.e. number of events  per interval of SNR in
logarithmic binning $d(\lg{S})$.  They found power-law segments in the
distributions with the slopes $-$2.3 and $-$2.9 at 0.43 and 8.8~\mbox{GHz},
respectively.  To compare this result with our study of the CPD
functions one has to convert the number of events presented by
\citet{cordes2004} from logarithmic binning to linear binning
$d(\lg{S})=\ln{10}dS/S$.  Then, for a power-law fit with linear binning
the slope will change from $-\beta$ to $-(\beta+1)$:
$$N(S)\propto (S)^{-\beta}d(\lg{S})\propto (S)^{-(\beta+1)}dS$$
for the power-law fit of the CPD $\gamma=\beta-1$, as was explained in
Sect. \ref{res}. Thus, the values of the power-law exponent $\beta$
found by \citet{cordes2004} in their histograms may by immediately
compared with our values of the exponents $\gamma$ for the power-law
fit of the CPDs.  In fact the inspection of the histogram of
\citet{cordes2004} for 0.43~GHz has enabled us to distinguish a
break at an $S/N$ value of about 30 where the slope changes from $-$2.3 to
$-$0.7 in the peak amplitude distribution of the main pulse GPs, while
the distribution for the interpulse GPs does not manifest such a
break. With the given equivalent flux system temperature of 1262~\mbox{Jy},
the receiver band of 12.5~\mbox{MHz}, and a sample interval of 128~$\mu s$, the
$S/N=30$ break point at 0.43~GHz will correspond to a pulse energy of
about 120000~$\mathrm{Jy}\cdot\mu \mathrm{s}$.\\
 In his PhD thesis \citet{moffett1997} presentes the CPD functions at
1.4~GHz based on observations conducted with the VLA with a time
resolution of 160~$\mu s$. In Fig. 4.3 (page 62) Moffett
distinguishes a break in power-law fit of the CPD for the main pulse GPs
at a level of 12~\mbox{Jy} corresponding to a pulse energy of about
2000~$\mathrm{Jy}\cdot\mu \mathrm{s}$, where the slope changes from $-$3.0 to $-$1.8 when
going from the high flux densities to the lower.  Again, the CPD for the
GPs originated at the longitudes of the interpulse (Figure 4.4) does
not show a break going straight with the slope of $-$1.7 until it merges
with noise at a level of about 6~\mbox{Jy}.\\
Comparing the values of break-point pulse energy (BPPE)
120000~$\mathrm{Jy}\cdot\mu \mathrm{s}$ at a frequency of 0.43 GHz (Cordes \etal
(2004)), 5000~$\mathrm{Jy}\cdot\mu \mathrm{s}$ at a frequency of 1200~\mbox{MHz} (this paper),
and 2000~$\mathrm{Jy}\cdot\mu \mathrm{s}$ at a frequency of 1400~\mbox{MHz} (Moffett (1997)),
we have found that the BPPE follows a simple power-law frequency
dependence $BPPE(\nu)\approx 7(\nu)^{-3.4} \mathrm{kJy}\cdot\mu \mathrm{s}$, with the
exponent $-$3.4 close to the mean spectral index for the main pulse
component of the average profile $-$3.0 \citep{moffett1997}.  The
conclusion can be considered as support for the suggestion that all
emission in the main component consists entirely of giant pulses
\citep{popov2006}.\\
\citet{lundgren1995} has collected about 30000 GPs from the Crab pulsar at
812 \mbox{MHz} with the Green Bank 43-m radio telescope in 10 days of
observations simultaneous with the Compton Gamma Ray Observatory
(CGRO) in May 1991. They did not distinguish GPs belonging to the main
pulse and the interpulse.  Their flux-density distribution was fitted
with a power-law function for $S > 200$~\mbox{Jy}, and the exponent was found
to be equal to $-$3.46, the exponent being equal to $-$2.46 for the
CPD used in our analysis.  The
BPPE value is expected to be 14~$\mathrm{kJy}\cdot\mu \mathrm{s}$ at 812 \mbox{MHz} according
to the frequency relation we derive above. The value corresponds to
50~\mbox{Jy} flux density with 307.5~$\mu s$ averaging time used by
\citep{lundgren1995}, and it is well below their threshold of
120~\mbox{Jy}. The dramatic roll-off at the rather high pulse energy of
about 200~$Jy\times 300\mu s= 60 \mathrm{kJy}\cdot\mu \mathrm{s}$ found by
\citep{lundgren1995} was not observed in the CPDs at 430, 1200, and
1400~\mbox{MHz} discussed above. It was not observed in the CPD at 600~\mbox{MHz}
either, which goes straight with the slope of $-$2.2 at least down to
a pulse energy of about 30~$\mathrm{kJy}\cdot\mu \mathrm{s}$ \citep{popov2006}.
Therefore, it is difficult to reconcile their results with those
from the many independent data sets mentioned above.\\
The power-law form of the observed CPDs can be compared with the field
statistics of possible emission mechanisms responsible for the
generation of GPs, as discussed by \citet{cairns2004} who used
the values of the exponents of PD functions for comparison.  The very
short-time flux  density variations are of particular interest, since they are
closely tied to the physics of the emission process. The power-law
indices $\gamma$ of the CPDs at high energies for the shortest GPs are
similar enough for both the Crab pulsar and the PSR B1937+21,
and the values were found to be in the range 1.4 to
1.8. Converting to indices of the probability distribution function
(PD) $\beta = \gamma +1$ gives the range 2.4 -- 2.8 for the values of
$\beta$. According to the normalization conditions used by Cairns, the
exponent $\alpha$ of the PD of the field
($P(\mathcal E) \propto \mathcal E^{-\alpha}$) is connected with the
exponent $\beta$ of the observed PD power-law fit by the relation
$\alpha=2\beta-1$, giving us the range of $\alpha$ from 3.8 to 4.6 to be
compared with the theoretical predictions of the field statistics
determined by the emission mechanism and propagation effects. The
observed breaks in the slope of the CPD functions at certain energies
have to be included in theoretical explanations, giving us an extra
parameter to constrain the source physics and emission mechanism.\\
The break in the slope of the CPD functions has important consequences
for the estimation of the total rate of GP generation.  \citet{popov2006}
made such an estimation for the Crab pulsar under the suggestion that
radio emission in the main pulse and in the interpulse consists
entirely of giant pulses. They found that about 10 giant pulses are
generated during one rotation period of the neutron star. In the
estimation they considered that pulse energies of the GPs follow a
power-law function with the exponent of $-$2.2 down to the threshold of
about 100~\mbox{Jy} in the peak flux density at a frequency of 600~\mbox{MHz},
and with the threshold considered as a real lower limit equivalent to the minimum
pulse energy of about 5000~$\mathrm{Jy}\cdot\mu \mathrm{s}$.  With their threshold of
about 20000~$\mathrm{Jy}\cdot\mu \mathrm{s}$ in GP detection, they did not notice the
break in the CPD slope. The break from $-$2.2 at high energies to $-$1.2
at low energies will notably change the estimate of the lower limit
for GP energies to about 1000~$\mathrm{Jy}\cdot\mu \mathrm{s}$, and the rate of GP
generation will increase at least by a factor of 2. The lower energy
limit for GP generation, if it exists, will serve as a crucial
constraint on the physics of the emission mechanism. To solve the
problem, it is necessary to test the low-intensity portion of the CPD
function with better sensitivity.  Such a study would be possible
using VLA or GMRT observations in phased array mode, thereby,
significantly reducing the impact from the Crab Nebula. A notable increase in
the recording band from 10 to 160~\mbox{MHz} recently achieved for
the PuMa II recording systems also makes new observations with the WSRT
very promising.\\
Finally, we summarize the main results of our analysis:\\
1) The CPDs were found
to be notably different for the GPs detected at the longitudes of the
main pulse and the interpulse.  We suppose that
the difference can be explained by the simple attenuation caused by a
beaming factor.\\
2) For the main pulse longitudes, the CPD are different for the GPs of
different effective widths with breaks in the CPD power-law indices
indicating steepening at high energies (see Table 1).\\
3) The CPD power-law indices ($\gamma_{high} \approx 1.7$)
for the group of short GPs for the Crab pulsar ($W_e<10\mu s$)
are  close to the  value
observed for the millisecond pulsar B1937+21, which
seems to generate only the shortest GPs;\\
4) GPs with a stronger peak
flux density were found to be of shorter duration.\\
The last of these
properties testifies in favor of nonlinear temporal models that
suggest that the higher the intensity, the narrower the pulse
width. Such an emission model was considered, for example, by
\citet{mikhailovskii1985}, who treated the micropulses as solitons of
the radio-wave envelope propagating through the magnetospheric plasma
of the pulsar.\\

\begin{acknowledgements}

 This investigation was supported in
part by the Russian Foundation for Fundamental Research (project number
04-02-16384). \\
                
\end{acknowledgements}

\bibliographystyle{aa}
\bibliography{biblio}

\begin{thebibliography}{33}
\expandafter\ifx\csname natexlab\endcsname\relax\def\natexlab#1{#1}\fi

\bibitem[{Allen(1973)}]{allen}
Allen, C.~W. 1973, Astrophysical quantities (University of London, Athlone
  Press)

\bibitem[{Argyle \& Gower(1972)}]{argyle1972}
Argyle, E. \& Gower, J. F.~R. 1972, ApJ, 175, L89

\bibitem[{Backer(1971)}]{backer1971}
Backer, D.~C. 1971, PhD thesis, Cornell Univ., Ithaca, NY

\bibitem[{Bietenholz {et~al.}(1997)Bietenholz, Kassim, Frail, Perley, Erickson,
  \& Hajian}]{bietenholz1997}
Bietenholz, M.~F., Kassim, N., Frail, D.~A., {et~al.} 1997, ApJ, 490, 291

\bibitem[{Cairns(2004)}]{cairns2004}
Cairns, I.~H. 2004, ApJ, 610, 948

\bibitem[{Cognard {et~al.}(1996)Cognard, Shrauner, Taylor, \&
  Thorsett}]{cognard1996}
Cognard, I., Shrauner, J.~A., Taylor, J.~H., \& Thorsett, S.~E. 1996, ApJ, 457,
  L81

\bibitem[{Cordes {et~al.}(2004)Cordes, Bhat, Hankins, McLaughlin, \&
  Kern}]{cordes2004}
Cordes, J.~M., Bhat, N. D.~R., Hankins, T.~H., McLaughlin, M.~A., \& Kern, J.
  2004, ApJ, 612, 375

\bibitem[{Cusumano {et~al.}(2003)Cusumano, Hermsen, Kramer, Kuiper,
  L{\"{o}}hmer, Massaro, Mineo, Nicastro, \& Stappers}]{cusumano2003}
Cusumano, G., Hermsen, W., Kramer, M., {et~al.} 2003, A\&A, 410, L9

\bibitem[{Ershov \& Kuzmin(2003)}]{ershov2003}
Ershov, A.~A. \& Kuzmin, A.~D. 2003, AstL, 29, 91, transl. from: PAZh, 2003,
  29, 111

\bibitem[{Hankins(1971)}]{hankins1971}
Hankins, T.~H. 1971, ApJ, 169, 487

\bibitem[{Hankins(2000)}]{hankins2000}
Hankins, T.~H. 2000, in ASP Conf. Ser. 202, Pulsar Astronomy -- 2000 and
  beyond, ed. M.~Kramer, N.~Wex, \& R.~Wielebinski (San Francisco: ASP), 165

\bibitem[{Hankins {et~al.}(2003)Hankins, Kern, Weatherall, \&
  Eilek}]{hankins2003}
Hankins, T.~H., Kern, J.~S., Weatherall, J.~C., \& Eilek, J.~A. 2003, Nature,
  422, 141

\bibitem[{Hankins \& Rickett(1975)}]{hankins1975}
Hankins, T.~H. \& Rickett, B.~J. 1975, in Methods in Computational Physics:
  Advances in Research and Applications, Vol. 14: Radio Astronomy, ed.
  B.~Alder, S.~Fernbach, \& M.~Rotenberg (New York: Academic Press, Inc.), 55

\bibitem[{Hesse \& Wielebinski(1974)}]{hess1974}
Hesse, K.~H. \& Wielebinski, R. 1974, A\&A, 31, 409

\bibitem[{Jessner {et~al.}(2005)Jessner, Slowikowska, Klein, Lesh, Jaroshek,
  Kanbach, \& Hankins}]{jessner2005}
Jessner, A., Slowikowska, A., Klein, B., {et~al.} 2005, AdSpR, 35, 1166

\bibitem[{Johnston \& Romani(2002)}]{johnston2002}
Johnston, S. \& Romani, R.~W. 2002, MNRAS, 332, 109

\bibitem[{Johnston {et~al.}(2004)Johnston, Romani, Roger, Marshall, \&
  Zhang}]{johnston2004}
Johnston, S., Romani, R.~W., Roger, W., Marshall, F.~E., \& Zhang, W. 2004,
  MNRAS, 355, 31

\bibitem[{Johnston {et~al.}(2001)Johnston, {Van Straten}, Kramer, \&
  Bailes}]{johnston2001}
Johnston, S., {Van Straten}, W., Kramer, M., \& Bailes, M. 2001, ApJ, 549, L101

\bibitem[{Kinkhabwala \& Thorsett(2000)}]{kink2000}
Kinkhabwala, A. \& Thorsett, S.~E. 2000, ApJ, 535, 365

\bibitem[{Knight {et~al.}(2006)Knight, Bailes, Manchester, Ord, \&
  Jacoby}]{knight2006}
Knight, H.~S., Bailes, M., Manchester, R.~N., Ord, S.~M., \& Jacoby, B.~A.
  2006, ApJ, 640, 941

\bibitem[{Kramer {et~al.}(2002)Kramer, Johnston, \& {Van Straten}}]{kramer2002}
Kramer, M., Johnston, S., \& {Van Straten}, W. 2002, MNRAS, 334, 523

\bibitem[{Kuzmin {et~al.}(2004)Kuzmin, Ershov, \& Losovsky}]{kuzmin2004}
Kuzmin, A.~D., Ershov, A.~A., \& Losovsky, B.~Y. 2004, AstL, 30, 247, transl.
  from: PAZh, 2004, 30, 285

\bibitem[{Kuzmin {et~al.}(2002)Kuzmin, {Kondrat'ev}, Kostyuk, Losovsky, Popov,
  Soglasnov, {D'Amico}, \& Montebugnoli}]{kuzmin2002}
Kuzmin, A.~D., {Kondrat'ev}, V.~I., Kostyuk, S.~V., {et~al.} 2002, AstL, 28,
  251, transl. from: PAZh, 2004, 28, 292

\bibitem[{Lundgren {et~al.}(1995)Lundgren, Cordes, Ulmer, Matz, Lomatch,
  Foster, \& Hankins}]{lundgren1995}
Lundgren, S.~C., Cordes, J.~M., Ulmer, M., {et~al.} 1995, ApJ, 453, 433

\bibitem[{Lyne(1982)}]{lyne1982}
Lyne, A.~G. 1982, Jodrell Bank Crab Pulsar Monthly Ephemeris ({\tt
  http://www.jb.man.ac.uk/~pulsar/crab.html})

\bibitem[{Mikhailovskii {et~al.}(1985)Mikhailovskii, Onishchenko, \&
  Smolyakov}]{mikhailovskii1985}
Mikhailovskii, A.~B., Onishchenko, O., \& Smolyakov, A.~I. 1985, SvAL, 11, 78,
  transl. from: PAZh, 1984, 11, 190

\bibitem[{Moffett(1997)}]{moffett1997}
Moffett, D.~A. 1997, PhD thesis, New Mexico Inst. Mining and Technology

\bibitem[{Moffett \& Hankins(1996)}]{moffett1996}
Moffett, D.~A. \& Hankins, T.~H. 1996, ApJ, 468, 779

\bibitem[{Popov {et~al.}(2006)Popov, Soglasnov, Kondratiev, Kostyuk, Ilyasov,
  \& Oreshko}]{popov2006}
Popov, M.~V., Soglasnov, V.~A., Kondratiev, V.~I., {et~al.} 2006, ARep, 83,
  660, transl. from: AZh, 2006, 83, 717

\bibitem[{Ritchings(1976)}]{ritchings1976}
Ritchings, R.~T. 1976, MNRAS, 176, 249

\bibitem[{Romani \& Johnston(2001)}]{romani2001}
Romani, R.~W. \& Johnston, S. 2001, ApJ, 557, L93

\bibitem[{Soglasnov {et~al.}(2004)Soglasnov, Popov, Bartel, Cannon, Novikov,
  Kondratiev, \& Altunin}]{soglasnov2004}
Soglasnov, V.~A., Popov, M.~V., Bartel, N., {et~al.} 2004, ApJ, 616, 439

\bibitem[{Voute {et~al.}(2002)Voute, Kouwenhoven, Langerak, Stappers, Driesens,
  Ramachandran, \& Beijard}]{voute2002}
Voute, J. L.~L., Kouwenhoven, M. L.~A., Langerak, J.~J., {et~al.} 2002, A\&A,
  385, 733

\end{thebibliography}

\end{document}